\documentclass{article}
\usepackage{ijcai25}

\usepackage{times}
\usepackage{soul}
\usepackage{url}
\usepackage[hidelinks]{hyperref}
\usepackage[utf8]{inputenc}
\usepackage[small]{caption}
\usepackage{graphicx}
\usepackage{amsmath}
\usepackage{amsthm}
\usepackage{booktabs}
\usepackage{algorithm}
\usepackage{algorithmic}
\usepackage[switch]{lineno}

\usepackage{graphicx}
\usepackage{subcaption}

\title{Can AI Explanations Make You Change Your Mind?}

\author{
Laura Spillner$^1$
\and
Rachel Ringe$^1$\and
Robert Porzel$^1$\And
Rainer Malaka$^1$\\
\affiliations
$^1$Digital Media Lab\\
University of Bremen
\emails
\{l.spillner, rringe, porzel, malaka\}@uni-bremen.de
}

\begin{document}

\maketitle

\begin{abstract}
    In the context of AI-based decision support systems, explanations can help users to judge when to trust the AI's suggestion, and when to question it. In this way, human oversight can prevent AI errors and biased decision-making. However, this rests on the assumption that users will consider explanations in enough detail to be able to catch such errors. We conducted an online study on trust in explainable DSS, and were surprised to find that in many cases, participants spent little time on the explanation and did not always consider it in detail. We present an exploratory analysis of this data, investigating what factors impact how carefully study participants consider AI explanations, and how this in turn impacts whether they are open to changing their mind based on what the AI suggests.
\end{abstract}

\section{Introduction}

This paper describes our quest to understand what factors influence participants' engagement with AI explanations, in a user study on decision support systems (DSS).  Originally, we wanted to investigate how the design of the interaction and the presentation of explanations impact users' trust in the system. We had developed several hypotheses, and designed two rounds of studies with which we hoped to answer our research questions. When we analyzed the data, however, we realized that we were operating under an incorrect assumption: We had assumed that our participants would look at and read all AI explanations in detail. We also assumed that it was in fact necessary to read explanations in detail in order to gain helpful information from them. The data we collected disproved the first assumption, and led us to question the second. 

When DSS systems are built for human-AI-collaboration, and humans have final say on a decision, the hope is that human oversight can prevent AI mistakes and AI bias \cite{lai_human_2019,alizadeh_user-friendly_2022}. The human-AI-team should ideally perform better than either human or AI can on their own. If that is not the case, the human should at least be able to prevent cases of disciminatory bias. Having the AI explain its decision suggestions is supposed to aid in this process \cite{chen_machine_2023}: It should help human users to realize when the AI would make a decision for discriminatory reasons. It could also enable humans to realize when there is an obvious error in the AI model, and potentially even find the reason for that error. Finally, good explanations would also help human users to understand the AI's reasoning better when it is correct, improving their understanding of the decision process, and making the reason for a decision transparent and understandable \cite{adadi_peeking_2018}. 

We have conducted an exploratory analysis of the data we collected, the results of which are presented below. With this analysis, we hoped to understand better what happened in our studies, and if and how participants considered the AI explanations. Which factors impacted in how much detail participants read the AI explanations? Which factors impacted whether or not they changed their mind based on the AI suggestions? Did the explanations still help them to understand when the AI was right, and when it erred? Based on this analysis, we discuss questions for future research directions and study design for explainable AI in the context of DSS.  

\section{Related Work}

One of the goals of explainable AI (XAI) is to increase trust in AI \cite{abdul_trends_2018}, and increase the trustworthiness of AI both for users and for people affected by its decisions \cite{barredo_arrieta_explainable_2020}. Thus, AI should be able to justify its decisions with user-comprehensible explanations to prevent mistakes \cite{adadi_peeking_2018,barredo_arrieta_explainable_2020}. Trust in an AI system is considered \emph{warranted} only when the AI is actually trustworthy \cite{jacovi_formalizing_2021}, and users should be able to judge when they should or should not follow the AI suggestion instead of blindly trusting its recommendations. 

With DSS, explanations for individual decisions are often not very interactive: the `correct' decision as predicted by the AI is presented together with a local explanation of this prediction \cite{liao_questioning_2020,lai_human_2019}. Users then have to make the final decision, providing human oversight and hopefully catching cases of AI error or bias \cite{chen_machine_2023}. Many have argued that AI explanations could better aid decision-making if they were more interactive or more conversational \cite{miller_explainable_2017,norkute_ai_2020,feldhus_mediators_2022,wiegreffe_reframing_2022,bertrand2023selective,battefeld2024ascodi,sovrano_explanatory_2024}. However, as explainable AI systems are being deployed, it is still important to understand how static explanations impact decision-making, and how their presentation can be fine-tuned to achieve the best possible outcome \cite{hoffman_metrics_2018,chromik2021human}. This means reducing both \emph{undertrust} (users being too skeptical of the AI) and \emph{overtrust} (users trusting the AI when it misguides them towards a wrong decision) \cite{jacovi_formalizing_2021}. 

Among previous works on XAI in a DSS context, some studies use a decision workflow in which the participant is shown the AI's suggestion immediately \cite{sivaraman2023ignore,chiang2023two,gaube2021ai,alufaisan2021does} when seeing a decision task, which we call a \emph{one-step workflow}. Other studies have the participant make an intuitive decision first and only then present the AI's recommendation \cite{he2023knowing,wang2021explanations}, which we call a \emph{two-step workflow}. How the choice of workflow impacts decision-making is not entirely clear: Fogliato et al. found that when given an AI recommendation and a confidence score, participants in the two-step workflow group were less likely to agree with the AI system \cite{fogliato2022}. Green and Chen conducted a large study on ethical decision-making, where one of their chosen conditions used a two-step workflow, while participants in another condition were provided with an explanation. They found that a two-step workflow improves performance of the human-AI-team \cite{green_2019}.
Bu\c{c}inca et al. conducted a study to determine if cognitive forcing designs can be used to combat overreliance on AI systems (overtrust) and compared three possible designs—one of which was a two-step workflow—to a simple XAI approach (one-step and static). Their results show that while cognitive forcing designs reduce overreliance more than the XAI approach, they were less liked by participants \cite{bucinca_2021}.

In addition, there have also been a number of studies on other methodological aspects in XAI research, such as the use of proxy tasks, and questionnaires to measure user trust. Bucina et al. found that the outcomes of studies using proxy tasks in XAI studies could not be used to predict the results of studies using the actual decision-making tasks. Moreover, they found that subjective measures could not be used as a predictor for performance on decision-making tasks \cite{bucina_proxytasks}. Byrne also cautions on the use of self-reported measures on trust and satisfaction to judge an explanation's quality in XAI studies, due to the illusion of explanatory depth—the tendency to overestimate one's understanding of a process or device \cite{byrne2023good}. A two-step workflow provides an alternative way to gauge users’ trust in the AI system: if the user’s first choice (before seeing the AI recommendation) is known, it can be measured how often they change their mind if the AI disagrees with their first choice, indicating that they trust the AI system enough to follow its recommendation over their own initial opinion. 

\section{Data Collection \& Analysis}

We conducted two rounds of an online decision task study that was designed to answer research questions on the effect of study design and the presentation of explanations on users’ trust in the AI system. In each case, participants had to make a binary decision, and were presented with a decision suggestion by an AI. In a first round, we tested the impact of the design of the decision workflow: one-step (participants were shown the AI suggestion immediately together with all the relevant data for the task) or two-step (they were first shown the data, had to make an initial choice, then saw the AI suggestion, and then had to log their final decision). We tested this with two versions of the AI system, one without explanation, and one that provided a local feature importance explanation by highlighting relevant data items in shades of red and green. In the second round, we repeated the two-step study with two more variants of explanation: a longer text describing the feature importance, and a bar chart showing the same information visually (also colored as in the highlighting). 

This paper does not present the results of these studies according to our original research questions. In analyzing the results, we were surprised that even though participants performed well on the tasks, and improved together with the AI, they often spent very little time considering the explanations and, in some cases, clearly did not look at them in detail, which went against our assumptions in designing the study. Thus, we present in this paper an exploratory analysis of the data collected, in hopes of understanding better how participants interact with AI explanations, and what factors actually impact whether AI can make them change their mind.

For this analysis, we only include the data from the two-step rounds, because this allows us to identify cases where they changed their mind (which we will call \emph{switch}), and how long they took for their own decision vs. for thinking about the AI suggestion. This data includes four explanation variants: no explanations, explanations in the form of highlighting on the data table, bar chart explanations, and full text explanations. All three types of explanations showed the same information (local feature importance) in different formats (see Appendix for screenshots).

\subsection{Study Design}

For the decision task, participants were asked to predict the likely academic career outcome of university students, with the stated goal of offering additional support to struggling students who needed it the most, because they were likely to otherwise drop out. They had to predict whether the student would complete their studies successfully or drop out (binary decision)—based on this, the additional help would be distributed. As basis we used an existing open source dataset on the academic career of Portuguese university students \cite{realinho2022predicting}\footnote{https://www.kaggle.com/datasets/thedevastator/higher-education-predictors-of-student-retention}. Participants saw data about each student such as their grades and number of credits received in the first two semesters, demographic and family background, and information about tuition and debt. 

\paragraph{Decision Task}

We had conducted a pre-study to assess the suitability of the use case: we wanted a task where both human participants and AI were reasonably good, but not perfect, at solving it. The goal of this was to have enough cases in the data in which either the participant or the AI were wrong to be able to analyze them. Humans in an in-person lab study solved a little less than 70\% of cases correctly, which we found to be a suitable baseline. Implementing an AI with a similar level of accuracy proved difficult, as simple models were able to solve the task with high accuracy (\textgreater95\%). Therefore, in the first round of studies, we utilized a kind of ``Wizard of Oz''-AI system, where we pre-defined what suggestion should be given (for an accuracy of 73\%, 11 out of 15 tasks correct). Then we calculated backwards for each case how well each feature fit the target prediction, to generate a post-hoc analogue to a feature importance explanation. For the second round of study, however, we updated the AI to a linear regression classifier model, to have a more realistic model. The model was artificially constrained to arrive at the same accuracy by limiting training data and iterations; explanations were calculated with the \texttt{SHAP} package in Python.

\paragraph{Online Study}

The study was conducted on the online platform Prolific, and participants were paid to take part in it. After a general introduction, they completed 15 decision tasks. In each, they were shown a table with data for a student from the dataset. They had to submit their first choice, then were shown the AI suggestion (and explanation), and then had to submit their final choice. We logged each choice as well as the decision times. Participants were not told if their decision was correct after each case, in order not to influence their trust in AI in this way. After completing all tasks, participants were asked to fill out a questionnaire. This included a modified version of the HCT questionnaire \cite{madsen2000measuring}, demographic information, and their familiarity with AI as well as the domain in question. 

\subsection{General Results}

There were 75 participants in each explanation condition, for 300 participants total that completed the two-step study setup (another 150 participants completed the one-step setup, but are not included in the following analysis). Some had to be excluded because they did not finish the study due to technical problems, or did not pass an attention test in the questionnaire (see below). 283 participants are included in the analysis below: 72 in the condition without any AI explanations, 74 with data highlighting explanations, 74 with full text explanations, and 63 with bar chart explanations. Participants were aged between 18 and 60 years, with an average age of 28 years (SD=9.8 years). 144 of the participants identified as female and 136 as male. None of the participants reported occupations that made them particular experts in the domain of student support, however, 108 of them reported they were students. For the final decisions, the mean decision time was 13.2s (SD=29.5, 95\% CI [12.2, 14.1]), median 5.9s. 


\paragraph{Attention Test}

In the second course of data collection, we added two attention tests based on recommendations for Prolific studies: one `standard' attention test in the final questionnaire (a question that told participants explicitly that it was a test and which answer to pick), as well as a self-designed `task-based' attention test with which we hoped to filter out participants who did not engage with the decision tasks seriously. For this, we hid the attention test in the explanation of the AI suggestion: it asked them to choose the option that was incorrect, thus did not align with the data, and also went against the AI suggestion.
8.7\% of participants did not pass the attention test in the questionnaire, and were excluded from continuing the study at that point. However, only 13.0\% of those who passed it also succeeded in the task-based attention test—87\% of participants failed that test. Our immediate reading of this result was that a) almost no participants had actually engaged with the tasks seriously, and b) because of this, we could not use the collected data to answer our research questions. 
It is, of course, possible that data from online data collection portals is simply not suitable for this kind of study. Not only did most participants not pass the attention test, but in 44.8\% of cases, they spent \textless 5s on the AI explanation. Still, participants were overall quite good at solving the task, in fact, they were about as good as the in-person participants in the lab pre-study. Moreover, there were extreme differences in how much time they took to consider the data and the AI suggestions (1s to several minutes). We very much wanted to understand whether participants had (usually) read the AI explanations or not, what impacted how much time they took to consider them, and their ultimate decision-making.

\subsection{Exploratory Data Analysis}

We decided to conduct an exploratory analysis of the data. We wanted to understand which factors impacted a) the time taken to consider the AI explanation before making a final decision and b) whether or not users would be convinced by the AI to change their mind compared to their initial choice, that is, switch their decision. All participants had solved multiple tasks, and each task was solved by multiple participants, meaning that the data is clustered and not independent. We were interested in both trial-level factors that differ for each task a participant solved (decision time, type of explanation, whether or not the AI suggestion matches the participant’s first choice), and participant-level traits that might modulate these (their familiarity with AI or the domain at hand).

\paragraph{Understanding Decision Time}

First, we employed a linear mixed-effects model to examine how these factors influence the amount of time taken to consider the second decision (after receiving an AI suggestion). The dependent variable (second decision time) was log-transformed to account for non-linearity because of the heavy right-skew (it is bounded at zero, and there were some large outliers, with some participants taking several minutes for some tasks). Fixed effects included: the time taken to decide on the first choice (log-transformed); the explanation type: none (default), feature importance conveyed through highlighting in the data table, through a bar chart, or explained in a text; whether first choice and AI suggestion differ (binary); whether participants had experience with AI (binary); and their perceived knowledge of the domain (Likert question, coded as -2 and +2). 
We included two-way interactions to test for moderation effects. To account for repeated measures, we also included random intercepts for participant and task. The model was implemented via the \texttt{lmerTest} package in R, with Satterthwaite-approximated degrees of freedom and p-values for fixed effects, and fit using restricted maximum likelihood. Because the model did not converge easily with the default optimization algorithm, we used the \emph{bobyqa} optimization algorithm with 100{,}000 iterations. With this setup, the model converged. Then, we validated model assumptions visually by plotting the residuals, confirming that the log transformation had improved homoscedasticity and reduced right skew.

\paragraph{Changing One's Mind}

To understand how the same factors influence switching behavior (binary dependent variable), we fit a generalized linear mixed-effects model. For this model, we only included in our analysis those cases where participants had good reason to switch to the opposite decision compared to their first choice, because the AI suggestion disagreed with their first choice. There were, in fact, a few cases where the AI suggestion matched the participant’s first choice, and they then switched to the opposite for their second choice—however, this only happened in 0.68\% of cases, thus it is difficult to analyze based on such few examples. 
Therefore, whether the AI suggestion matched was not one of the fixed effects in this model. Other than that, we included the same fixed effects as before, and in addition, the time to decide on the second choice (log-transformed). Again, we included two-way interactions to account for moderation effects, and random intercepts for participant and task. We implemented the generalized LMM in R with a binomial outcome (logit link) using the \texttt{lme4} package, and again used the \emph{bobyqa} optimizer with a maximum of 100{,}000 iterations.

\section{Results}

\subsection{Why did our participants fail the in-explanation attention test?}

On seeing that some participants had taken very little time for their first and/or second decision, we assumed some of them had not engaged with the task seriously: that they had clicked on either option without looking at the data or the AI suggestion in much detail. 
If we could filter out those people who had not seriously considered each task, we hoped that we would be left with a subset of participants based on whose data we could answer our research questions.

However, it turned out that our assumption was, again, wrong. First, we could not identify any obvious outliers on the lower end of the decision times. 
Next, we assumed that the non-attentive participants would obviously perform worse at the task, and thus checked how accuracy differed based on decision time. Across all trials, the participants picked the correct option immediately (before seeing the AI suggestion) 65.5\% of the time, and, after seeing the AI suggestion, solved 72.3\% of the cases correctly. Surprisingly, accuracy does not differ based on time: for the first choice, accuracy was: 65.4\% for \textless 2s, 66.8\% for 2s-5s, 63.4\% for 5s-20s, 64.6\% for \textgreater 20s. For the second and final choice, they picked the correct decision: 75.1\% for \textless 2s, 73.3\% for 2-5s, 72.7\% for 5-20s, and 66.8\% for \textgreater 20s. Clearly, even when participants only looked at the explanations for a few seconds, they still attempted to solve the task as seriously as those who took longer.

It stands to reason that many of our study participants were indeed able to solve many tasks successfully by only glancing at the data and the AI suggestion. Even though they did not catch the (in retrospect, too well hidden) attention test in the AI explanations, they must have gotten helpful information from the AI suggestions and the explanations in many other cases: We used McNemar's Test (for paired binary data) to compare their first choices (on their own, 65.5\% correct) with the final decisions (with the AI, 72.3\% correct). The test revealed a statistically significant difference, $\chi^2(1, N = 3544) = 91.18$, $p < .001$, indicating that accuracy together with the AI was significantly higher.



Based on these results, we concluded that the attention test we added was not well designed to measure whether participants engaged with the tasks seriously. It did, however, reveal something else: that the vast majority of participants did not read every explanation in detail. 87\% of the participants who saw the attention test failed it. In 100\% of the attention test trials, the AI suggestion was the same as the participant’s first choice (in this task, all participants’ first choice was the correct decision; it was indeed the easiest task of the set). Do participants only engage with explanations in detail when they hope to gain new information from them, such as when the AI disagrees with their initial choice? 

\begin{table}[]
\begin{tabular}{lllll}
\multicolumn{1}{l}{} & \textbf{b} & \textbf{exp(b)} & \textbf{95\% CI}                                       & \textbf{p}       \\ \hline
Intercept                          & 1.397               & 4.044               & 3.209-5.096                                         &  \\ \hline
\textit{Explanation}               &                     &                     &                                                        &                  \\
(highlight)                        & 0.625               & 1.868               & 1.359-2.567                                            & ***            \\
(text)                             & 1.381               & 3.979               & 2.705-5.854                                            & *** \\
(bar chart)                        & 1.152               & 3.164               & 2.053-4.876                                            & *** \\ \hline
\textit{AI agrees}       & -0.608              & 0.544               & 0.477-0.622                                            & *** \\ \hline
\textit{Time (log)}       & 0.234              & 1.263               & 1.225-1.303                                            & *** \\ \hline
\textit{Knowledge ...}                 &                     &                     &                                                        &                  \\
\textit{of AI}              & 0.017               & 1.017               & 0.796-1.299                                            &             \\
\textit{of domain}          & -0.066              & 0.937               & 0.836-1.049                                            &            
\end{tabular}
\caption{Results of the LMM for decision time. Effect (b) is on log scale, exp($b$) is the multiplicative effect on the time. Significance levels: * $p < .05$, ** $p < .01$, *** $p < .001$. Two-way interactions are omitted, as none of them had notable or significant effects.}
\label{tab:time}
\end{table}

\subsection{What impacts how much time participants take to engage with AI explanations?}

We hoped that the log-transformed LMM would help us to understand better which factors impact how much time participants took to consider their second choice—and thus understand in which cases participants do or do not look at explanations in detail, and what impacts whether or not they do. The model, with random intercepts for participants and tasks, explained some but not all variability in second decision times: The intercept ($b = 1.41$, SE $= 0.13$) corresponds to about 4.0 seconds. Random effect variances were $0.165$ (participants), $0.007$ (tasks), and $0.494$ for residual error. 
The marginal $R^2$ (variance explained by fixed effects) was 0.474, and the conditional $R^2$ (variance explained by both fixed and random effects) was 0.610. None of the participant-level traits directly significantly influenced the time for the second decision, or significantly modulated the effect of the trial-level factors. However, all trial-level fixed effects impacted the decision time, as can be seen in Table \ref{tab:time}. 


\begin{figure}
    \centering
    \includegraphics[width=0.9\linewidth]{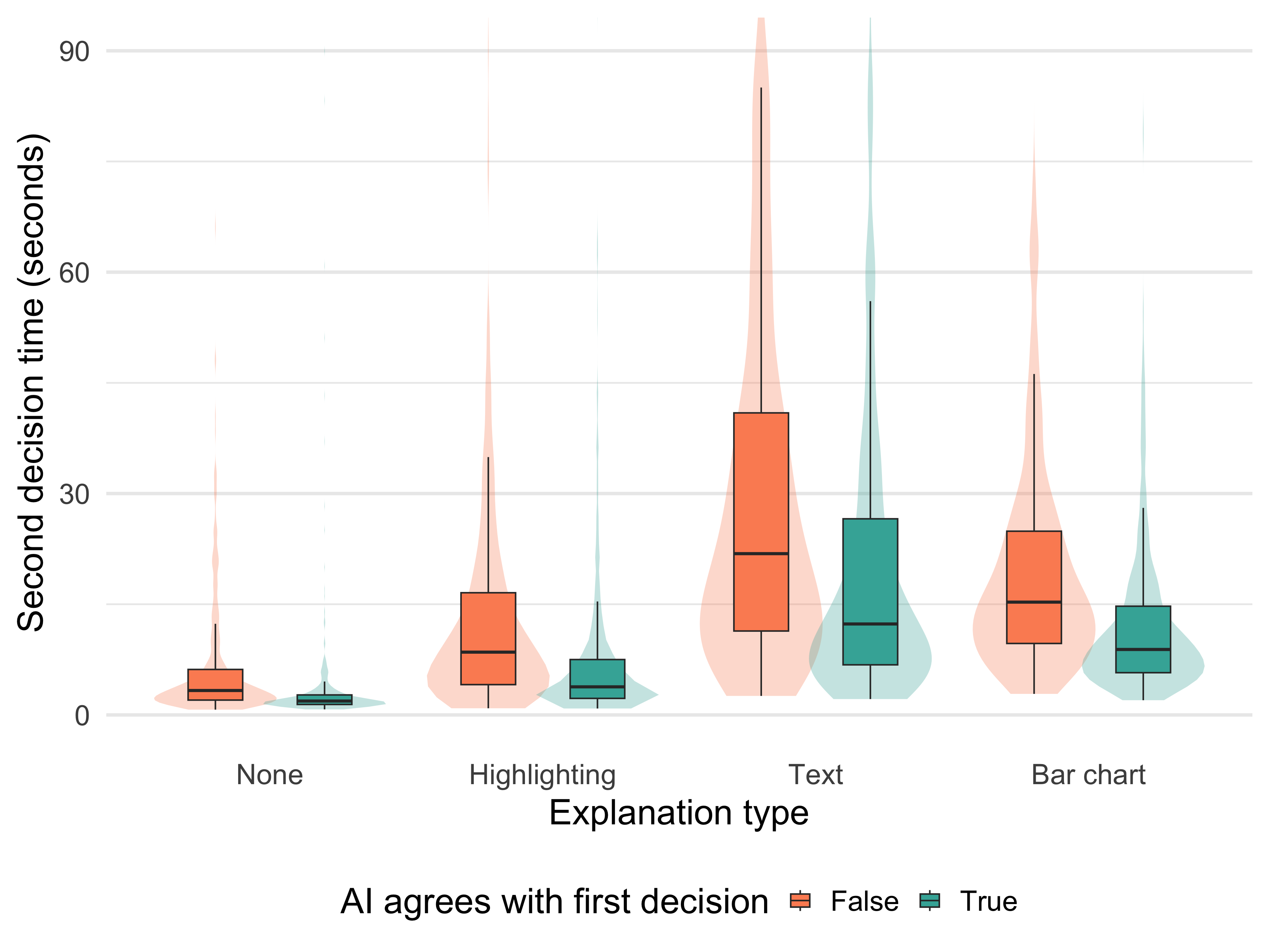}
    \caption{Time taken for second choice, by explanation and whether the AI suggestion was the same as the participant's first choice.}
    \label{fig:time-by-expl+AI}
\end{figure}

\paragraph{Time}

There is indeed a relationship between the two decision times: for every doubling of the time taken for the first decision, the time for the second decision increases by 26\%. This could indicate that when participants are less sure about the correct decision in a given task, they take more time to consider the AI suggestion and its explanation.

\paragraph{AI Suggestion}

Most strikingly, across all explanation types, participants took only about half as much time (54\%) for their second decision when the AI agreed with them compared to when it differed, as shown in Figure \ref{fig:time-by-expl+AI}. 

\paragraph{Explanation Type}

Explanation types also impacted how much time was taken (see Figure \ref{fig:time-by-expl+AI}). Compared to no explanations at all, highlighting of features according to their importance led to 1.9x the deliberation time, the bar chart illustrating feature importances to 3.2x, and the text describing the feature importances to 4.0x as much time. 


\subsection{What impacts whether or not participants change their mind?}

We fit a generalized LMM to analyze which factors impact the likelihood of switching. As described in Chapter 3.3, this only includes trials where the AI suggestion disagreed with participants' first choice. Overall, they switched their decision 42.9\% of the time in this case. Table \ref{tab:switch} shows the results. 


\paragraph{Decision Time}

Time does appear to have a small effect on switching likelihood, although it is only a significant difference for the time taken to deliberate on the first decision: when first decision time doubles, the likelihood to switch decreases by 18\%, while a doubling of the second decision time increases the likelihood of a switch by 13\%. 

\begin{table}[]
\begin{tabular}{lllll}
\multicolumn{1}{l}{} & \textbf{b} & \textbf{exp(b)} & \textbf{95\% CI}                                       & \textbf{p}       \\ \hline
Intercept                          & -0.555              & 0.574               & 3.164-5.266                                            &  \\ \hline
\textit{Explanation}               &                     &                     &                                                        &                  \\
(highlight)                        & 0.377               & 1.458               & 0.633-3.358                                            &                  \\
(text)                             & 1.878               & 6.541               & 1.974-21.675                                           & **               \\
(bar chart)                        & 1.597               & 4.938               & 1.176-20.727                                           & *                \\ \hline
\textit{Decision time}             &                     &                     &                                                        &                  \\
\textit{first (log)}               & -0.197              & 0.821               & 0.696-0.968                                            & *                \\
\textit{second (log)}              & 0.121               & 1.129               & 0.967-1.318                                            &                  \\ \hline
\textit{Knowledge}                 &                     &                     &                                                        &                  \\
of AI (True)                       & -0.229              & 0.795               & 0.421-1.502                                            &                  \\
of domain                          & 0.064               & 0.938               & 0.692-1.273                                            &                  \\ 
AI x text                          & -1.273              & 0.280               & 0.089-0.882                                            & *                \\
\end{tabular}
\caption{Results of the GLMM for switching decisions. Effect (b) is on log scale, exp($b$) is the multiplicative effect on the likelihood for switching. Significance levels: * $p < .05$, ** $p < .01$, *** $p < .001$. Two-way interactions, other than that of AI knowledge and text explanations, were not significant.}
\label{tab:switch}
\end{table}


\begin{figure*}[ht]
  \centering
  \begin{subfigure}[b]{0.33\textwidth}
    \includegraphics[width=\textwidth]{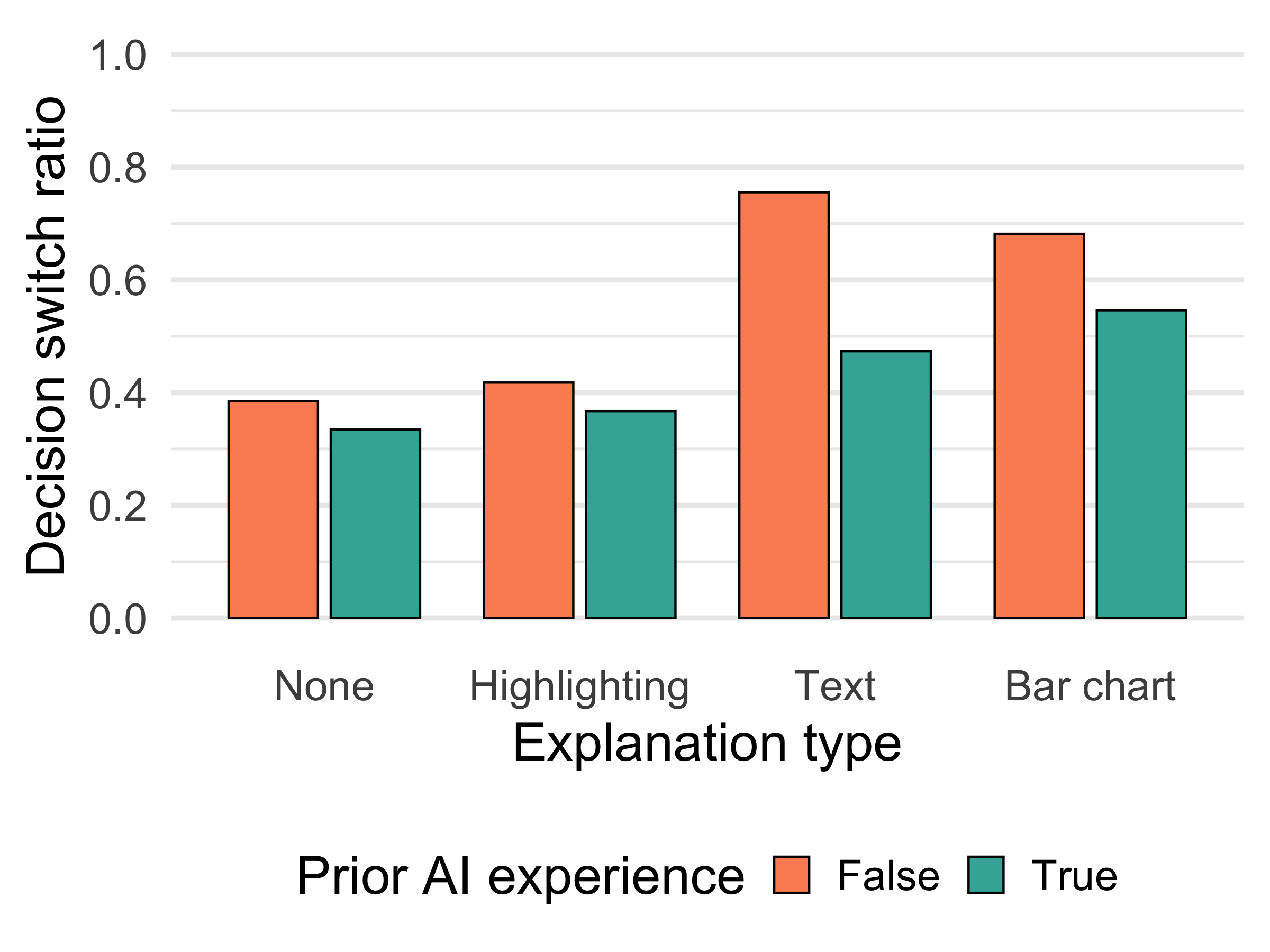}
    \caption{Switch ratios across all trials where AI \\ disagreed with participant's first choice.}
    \label{fig:switch-by-expl+AIknow}
  \end{subfigure}
  \hfill
  \begin{subfigure}[b]{0.33\textwidth}
    \includegraphics[width=\textwidth]{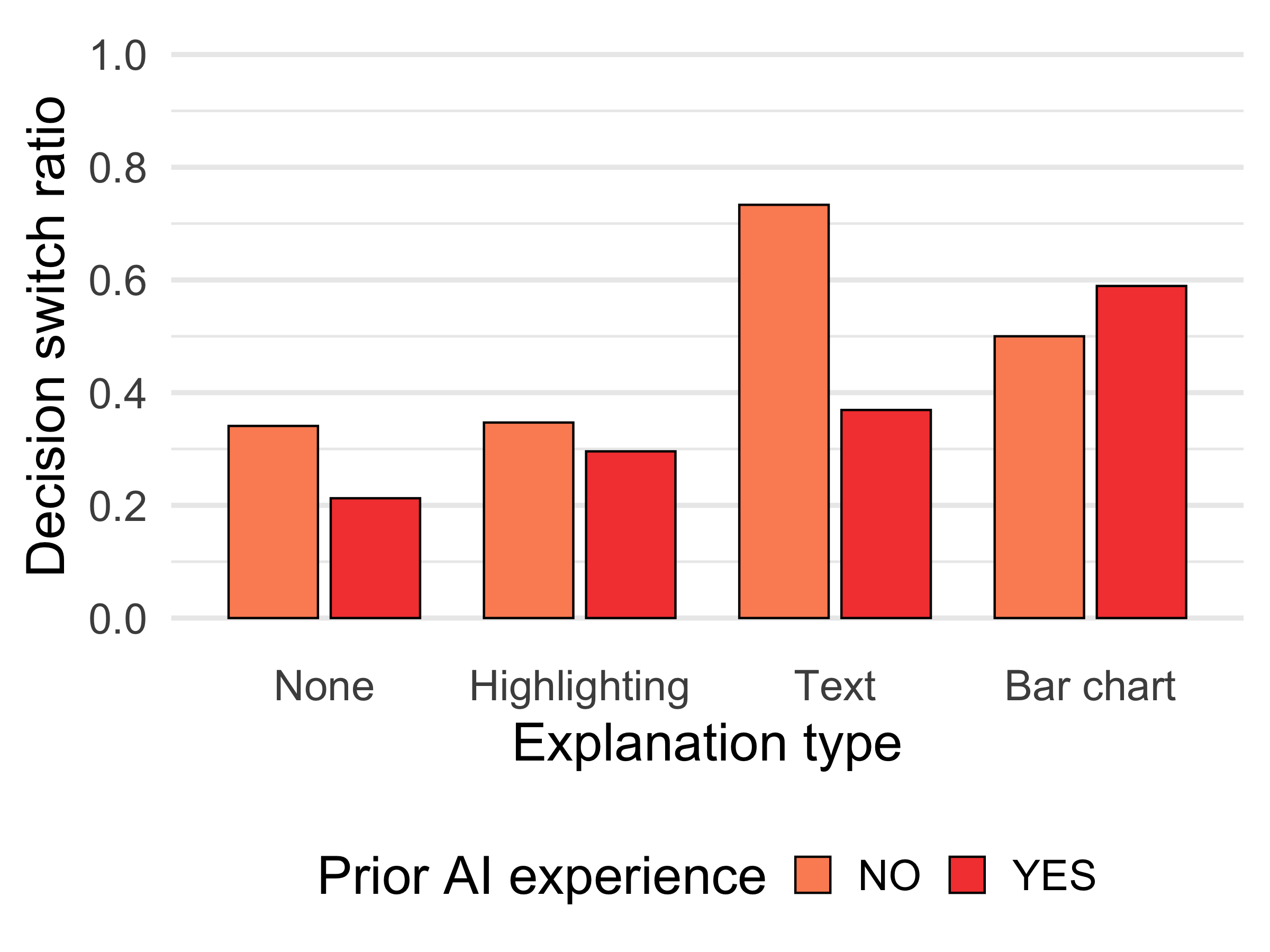}
    \caption{Switch ratios when first choice was correct, AI was wrong (switch = overtrust).}
    \label{fig:switch_bad-AI}
  \end{subfigure}
  \hfill
  \begin{subfigure}[b]{0.33\textwidth}
    \includegraphics[width=\textwidth]{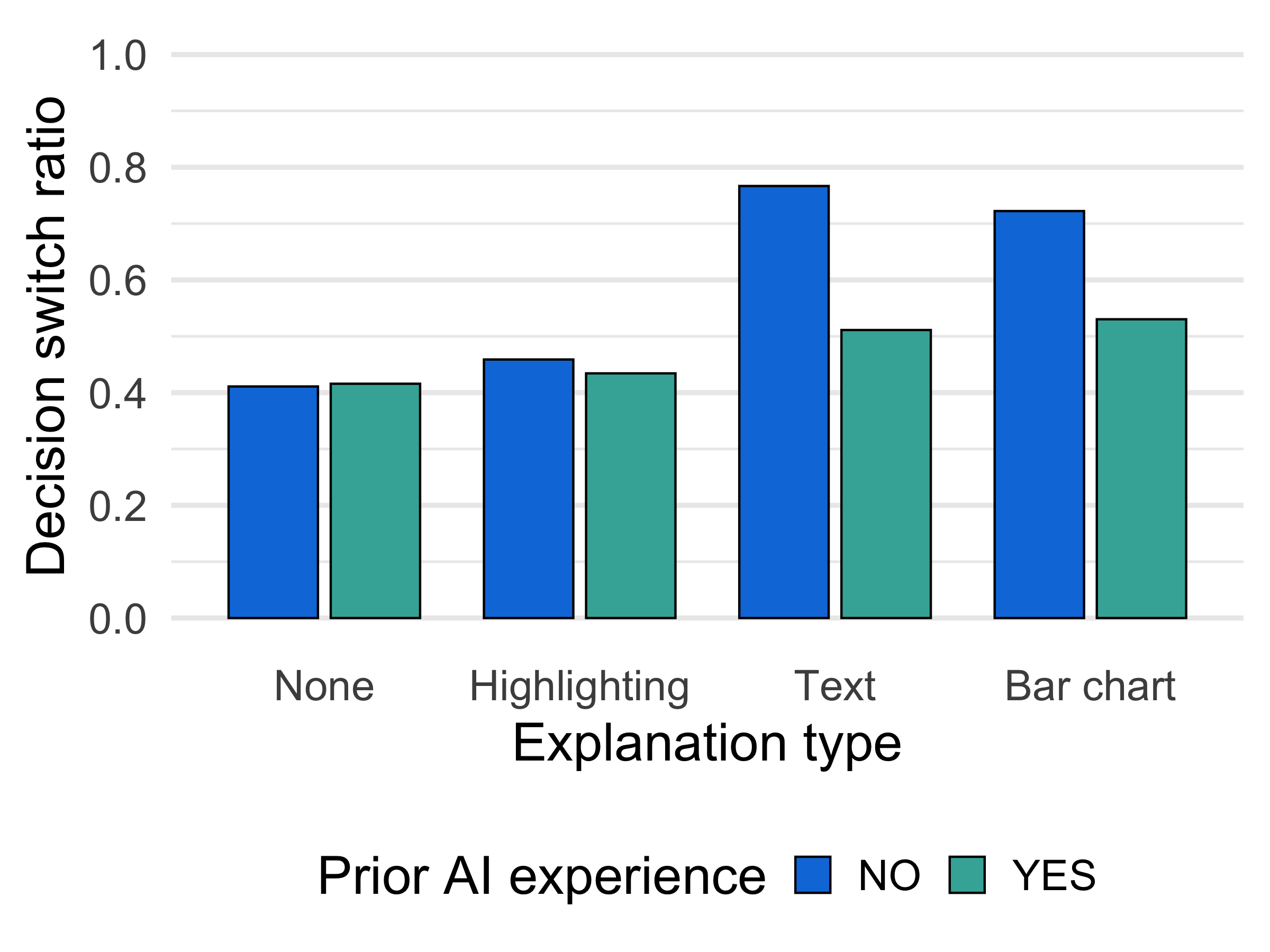}
    \caption{Switch ratios when first choice was wrong, AI was correct (no switch = undertrust).}
    \label{fig:switch_good-AI}
  \end{subfigure}
  \caption{Switch ratio by explanation and participants' AI experience. Only those trials where AI suggestion differed from user's first choice.}
  \label{fig:switch-by-AI+AIknow+expl}
\end{figure*}

\paragraph{Explanation Type}

Only bar chart and text explanation had a significant effect on participants’ likelihood to switch (these were also the two types that participants spent the most time on), as Figure \ref{fig:switch-by-expl+AIknow} illustrates. With bar chart explanations, participants were 4.9x as likely to switch compared to no explanation at all; with a text description of feature importance, they were 6.5x as likely to switch. The explanations contained the same content regardless of their presentation style. 

\paragraph{Participant Traits}

There is indeed some indication that participant-level traits modulate the effect that explanation type has on switching behavior. In particular, participants’ prior familiarity with AI significantly reduced the otherwise positive (in terms of making switching more likely) effect of text explanations, as Figure \ref{fig:switch-by-expl+AIknow} shows. 

\subsection{Did explanations help participants make correct decisions?}

In cases of disagreement, where the AI suggestion differs from the participant's first choice, a switch means that in their second choice, the participant followed the AI suggestion and reversed their initial choice. The switch ratio thus indicates the trust in the AI over one's own intuition. 
We wanted to gain some insight on possible differences between warranted trust in the AI (switching when one is wrong and the AI suggestion is correct), overtrust (switching when the AI is incorrect, thus being misled by the AI), and undertrust (sticking to ones initial, incorrect choice when the AI suggests the correct option). 
This was not part of the model, rather, we include here a visual/descriptive analysis. Figure \ref{fig:switch-by-expl+AIknow} shows the switch ratio across all combinations of explanation type and participants' AI experience. Next, we further divide this data between those cases where the AI was wrong (switching = overtrust) and those where the AI was right (switching = warranted trust, not switching = undertrust). 

Figure \ref{fig:switch-by-AI+AIknow+expl} shows this as two subfigures: in Figure \ref{fig:switch_bad-AI}, cases are shown in which the participant’s first choice was correct and the AI suggestion was wrong (higher red/orange bar means more overtrust). In Figure \ref{fig:switch_good-AI}, cases are shown in which the participant’s first choice was incorrect and the AI suggestion was right (higher blue/green bar means more warranted trust, lower bar means more undertrust). The `ideal' outcome would be to find that some explanations led to very low overtrust and high warranted trust for all or at least some participants. Overall, warranted trust was indeed higher than overtrust: A chi-squared test showed that the switch rate was significantly higher when the AI was right, with 48\% compared to 34\% when the AI was wrong ($\chi^2(1, N = 3544) = 36.42$, $p < .001$). However, not trusting the AI when it is right in half of all cases, and still trusting it when it is wrong in one third of cases, does not seem a very ideal scenario. 

\paragraph{Prior Experience with AI}

Across all cases of overtrust (Figure \ref{fig:switch_bad-AI}), participants who had prior experience with AI were less likely to be misled (32.3\% across all explanation types) than those without (40.2\%). Text explanations appear to be best at misleading those without AI experience into wrongly trusting the AI suggestion (73.3\%), while the bar charts mislead those with AI experience most often (59.0\%). 
Participants with AI experience also had higher rates of undertrust (Figure \ref{fig:switch_good-AI}): they only switched 47.1\% of the time when they were wrong and the AI was right, while those without prior experience with AI switched 51.0\% of the time. 

\paragraph{Explanation Type}

Texts and bar charts were more suited to convincing participants to trust the AI when it was right, compared to no or highlight explanations (Figure \ref{fig:switch_good-AI}), but the split based on AI experience reveals that this was only really the case for those without it (76.7\% and 72.6\%, respectively) than those with experience (51.1\% and 53.0\%, respectively, about the same as the average rate of warranted trust).
Independently of whether or not the AI suggestion was correct, text explanations seemed to be most convincing to participants without prior AI experience (see also Figure \ref{fig:switch-by-expl+AIknow}). As overtrust (73.3\%) is as high as warranted trust (76.7\%), it appears that they are not very good at helping these participants to differentiate whether or not the AI suggestion is likely to be correct based on the explanation. 
For those participants who reported having had prior experience with AI, text explanations perform best, as there are fewer instances of overtrust (36.9\%) and adequate levels of warranted trust (51.1\%). 

\section{Discussion and Conclusion}


\subsection{Insights from the Explorative Analysis}

\paragraph{Why did our participants fail the in-explanation attention test?}

It appears that the failed attention tests do not show that participants did not attempt the tasks seriously, but rather, that users of AI DSS likely do not read explanations in detail when the task is not difficult and the AI suggestion is not surprising—when it confirms their initial view on what would be the correct choice. 
We hypothesize that if they do not have reason to question the AI suggestion, they are not as interested in understanding how the AI suggestion came to be.

\paragraph{What impacts how much time participants take to engage with AI explanations?}

Participants spent much less time considering the AI suggestion and its explanation when it confirmed their initial choice, and also spent less time on the second choice if they had made the first one more quickly. Apart from this, explanation type significantly impacted how much time was taken to consider the explanations. 
We believe that much of this difference is likely due to the inherent differences in length/complexity of the explanation types. It takes considerably longer to simply read the text explanation compared to looking at either the figure or the highlighting. In the figure, moreover, one has to find those features that one is interested in, while the highlighting was added to the data table that participants were already familiar with. 
However, the time difference does align with how successful the different types of explanations are at convincing participants to disregard their first choice in favor of the AI suggestion.

\paragraph{What impacts whether or not participants change their mind?}

Those types of explanations that made participants engage with them for longer (texts and bar charts) were also more likely to make them change their mind. 
Does this actually translate to an improvement in understanding of the content of the explanation—that is, does it lead to warranted trust without increasing overtrust?
In general, the increase in trust overall is much stronger for those participants with no prior experience with AI. Those with experience remained more skeptical, and were not as easily swayed by text explanations. Are those who have experience with AI better at interpreting the explanations to make the correct decision, or are they simply more skeptical in general?
Furthermore, when participants had taken longer to consider their first choice, they were significantly less likely to change their mind—indicating that they became attached to the choice they had already thought about in detail. This is very relevant to the design of this kind of decision task: a two-step setup nudges users to spend time making up their own mind before seeing the AI suggestion. Does this only reduce overtrust (as suggested by \cite{bucinca_2021}, or might it also increase undertrust?

\paragraph{Did explanations help participants make correct decisions?}

Based on the visual inspection of the switch ratios across the different groups, we think that in this study, explanations could only have been helpful (in terms of preventing overtrust and facilitating warranted trust) to participants who had prior experience with AI systems. For those without experience, especially textual explanations were highly convincing, and led to as much unwanted overtrust as warranted trust. However, no explanation type appears to have been very helpful in this regard. Overall, overtrust happened in about a third of relevant cases, and undertrust in about half. The same rates, but no better, were achieved when text explanations were seen by AI-experienced participants, which was the most successful combination. All other subgroups either had low overtrust but even higher undertrust, or less undertrust but much higher overtrust.

\subsection{Questions for Future Work}

\paragraph{Attention to Explanations}

Our unwritten assumption when originally designing these studies was that participants would be motivated to look at all explanations with equal care. Thus, we assumed that they would be able to judge whether the AI explanation seemed reasonable, or whether it betrayed bias in the AI’s decision-making. The fact that participants often did not read the explanations in detail made us question how to design follow-up studies. Would users in a real-world scenario follow a similar pattern? Would they look at explanations in more detail when they expect to find new information in them, e.g., because the AI suggestion went counter to their instinctive choice? Can we actually test e.g. the effectiveness of full text vs. visual explanations if we do not know in which cases participants will consider the explanations carefully? What informs participants’ final decision: the presence of an explanation, its general format, some few important pieces of information, or the full content of the explanation—and how can we tell, without biasing the results because participants know we are recording the data?

\paragraph{Two-Step Study Design}

This explorative analysis is based on data where the participants first had to record their own, more or less instinctive, choice, and only then were able to see the AI suggestion. We do believe that the pattern we observed, that attention to the explanations depends on how surprising the AI explanation is, likely translates to other cases where this first choice is not explicitly recorded. It could be that there are some domains where (always depending also on the interface of the AI DSS) people are more likely to quickly form an instinctive opinion of what the right decision is upon looking at the data, whereas in others it takes more time to understand a case and make up your mind. In the former kind, the effect of differences in the explanation content could only truly be tested in those cases where the AI suggestion differed from the participants’ immediate, instinctive choice. Recording this choice, however, also forces participants to make a choice. Do we want users to make up their mind before seeing the AI suggestion, after having seen it, or perhaps arrive at a decision in some kind of collaborative process together with the AI? Would it be better not to let them know at all what the AI suggests? Such a system could make users explain their own decision, giving them the AI model as support to analyze how individual features might influence the final decision.

\paragraph{Measuring Attention to Explanations}

When and why do users engage with AI explanations? Which part of the explanation and its presentation impacts their final decision and their trust in the AI? Ultimately, we want to improve the presentation of explanations so that helpful information is easily available: the content of the explanation should aid users in deciding when to trust the AI over their instinct, and when to trust their own instinct over the AI prediction. The findings from this dataset, however, make us wonder how to measure any aspect of this without also biasing the outcome. How can we know if there was a ``first choice'', without forcing participants to make it? How can we find out in which cases users will be interested in an AI explanation in the real world? From our own experience (e.g., with large language models that provide sources or reasoning traces for their claims), one often takes into account the presence of a source, but not so much its actual content—might not the same be the case with DSS AI explanations? Can we measure what part of the explanation participants actually look at, without making them aware of the fact that we are doing it, e.g., by having them install eye-tracking software? How could we test their understanding of explanations,  without the testing itself nudging them to engage with the explanations in more detail?

\subsection{Conclusion}

When humans collaborate with AI systems, good explanations are supposed to help users evaluate the AI's suggestions. They should help them filter out biased or discriminatory AI decisions. If the user would have acted differently from what the AI suggests, explanations should help them decide whether to trust the AI, or trust their own intuition. 

We set out to conduct two studies designed to test hypotheses about the effect that the decision workflow, the presence of explanations, and their presentation as text or visual description, have on users' trust in AI DSS. Instead, we found that in many cases, participants only spent very little time considering the explanations. They often did not consider the content of the explanations in detail, especially when the AI suggestion matched what they already thought was the right decision. We had them submit their own opinion first before seeing the AI suggestion, and found that the more time they spent making up their own mind, the less likely they were to be convinced by the AI to change their mind. Participants who had prior experience with AI were especially skeptical and less willing to change their mind, while those without prior AI knowledge were easily swayed to trust a misleading AI suggestion over their initial, correct, choice. 

Based on this, we now have a wealth of new questions not just about the system design factors that impact users' decision-making together with AI, but also about the complexities of measuring their effects. In the future, we plan to follow two avenues of research: a) the meta-level at which study design impacts participants' attention to explanation, and b) alternative methods of presenting AI insight, e.g. by letting users choose whether to see explanations, letting them select which features should be explained, or designing the decision workflow as a collaborative process without explicit choices or suggestions in the beginning.










\bibliographystyle{named}
\bibliography{ijcai25}

\cleardoublepage

\onecolumn

\appendix

\section*{Appendix}

\begin{figure}[h!]
    \centering
    \includegraphics[width=0.5\linewidth]{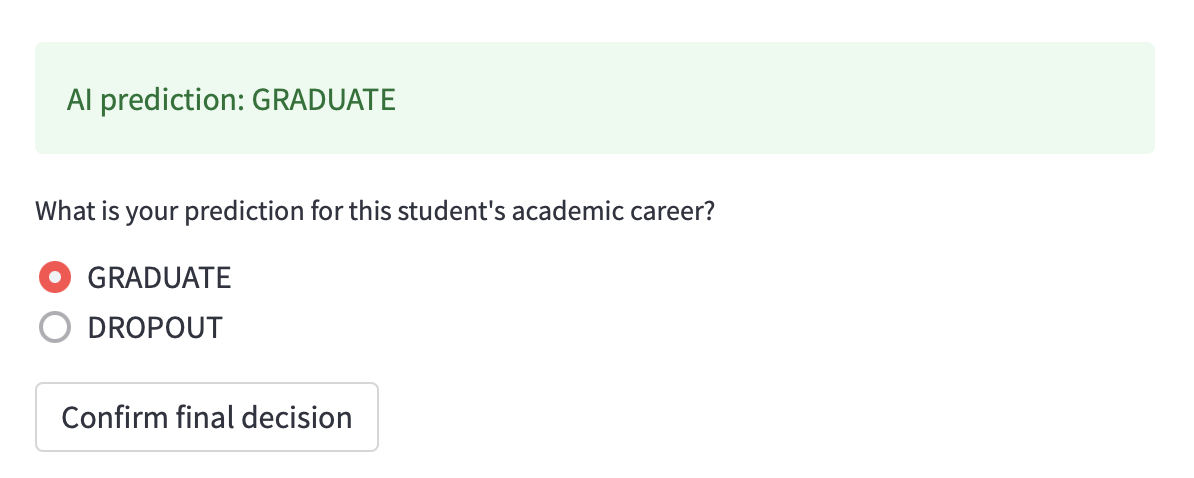}
    \caption{No explanation: The AI decision suggestion is provided without any explanation.}
    \label{fig:X_no}
\end{figure}

\begin{figure}[h!]
    \centering
    \includegraphics[width=0.8\linewidth]{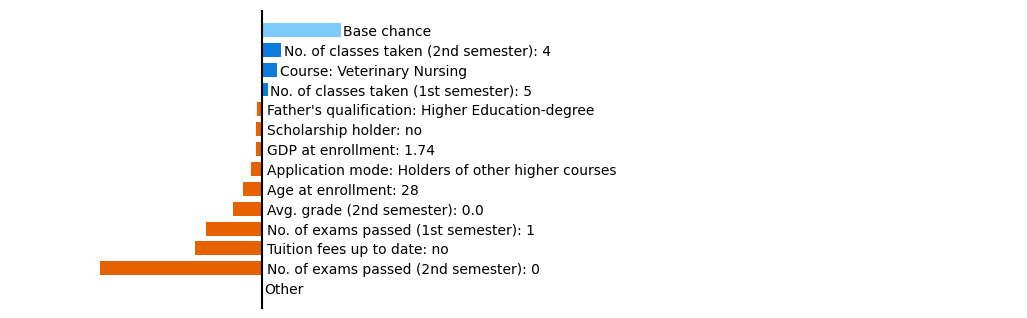}
    \caption{Bar Chart: Feature importance is visualized as a bar chart showing the respective (positive or negative) contributions of the most relevant features compared to the base chance of graduation. Participants were shown an example and given an explanation of how to read the chart in the tutorial, and the explanation is shown in a collapsible UI element beside each AI explanation.}
    \label{fig:X_vis}
\end{figure}

\begin{figure}[h!]
    \centering
    \includegraphics[width=0.9\linewidth]{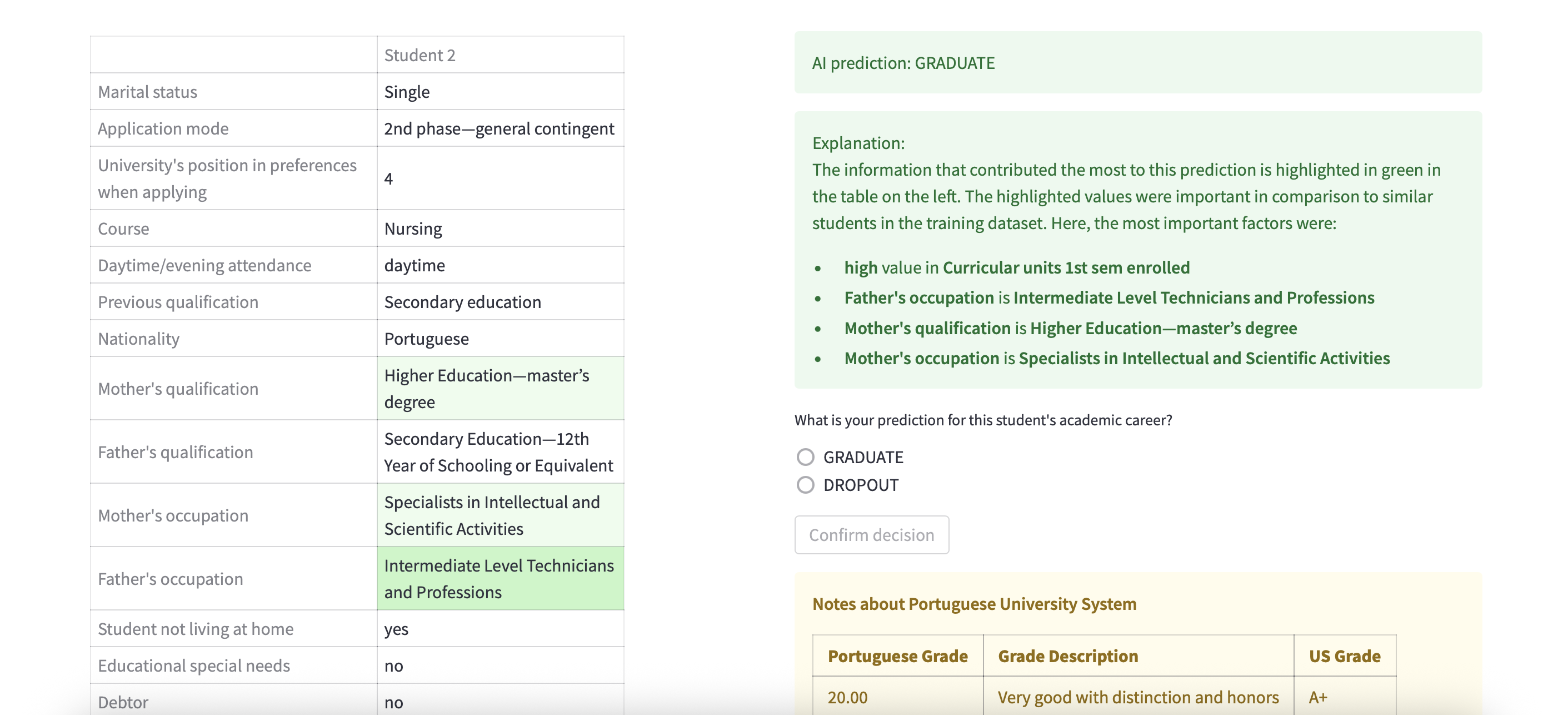}
    \caption{Highlighting: To explain feature importance, those features that contributed (in this case positively) to the decision are highlighted in the student information table in green (positive) or red (negative), with varying degrees of transparency indicating the strength of the contribution. Participants were given an explanation of how to understand feature importance in the tutorial, and a short explanation is provided next to the table.}
    \label{fig:X_high}
\end{figure}

\begin{figure*}
    \centering
    \includegraphics[width=0.8\linewidth]{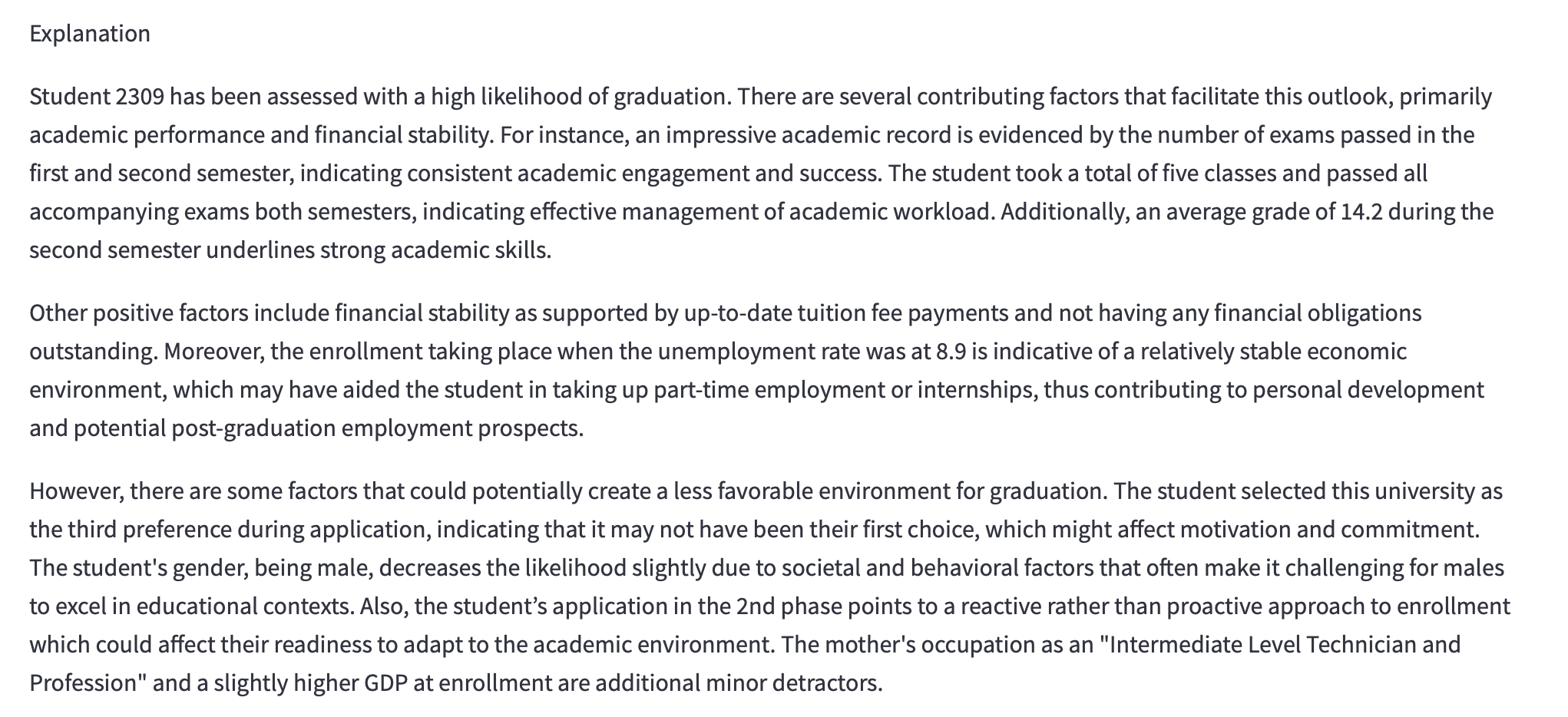}
    \caption{Text: The positive or negative influence of the most important features on the AI decision is provided in text form, giving possible reasons for why certain features would have positive or negative impact. The text explanations were generated with ChatGPT based on the feature importance values also used for visual explanations.}
    \label{fig:X_text}
\end{figure*}

\end{document}